\documentclass[final,twocolumn,showpacs,preprintnumbers,amsmath,amssymb,superscriptaddress]{revtex4}
\usepackage[utf8]{inputenc}
\usepackage{epsfig}
\usepackage{epstopdf}
\usepackage{amsmath,amssymb,amsthm}
\usepackage{dcolumn}
\usepackage{bm}
\usepackage{graphicx}
\usepackage{subfig}
\usepackage[justification=RaggedRight]{caption}
\usepackage[]{natbib}
\usepackage{hyperref}

\begin{document}
	
	\title{$N$-Block Separable Random Phase Approximation: \\Application to metal clusters and $C_{60}$ fullerene} 
	\author{D. I. Palade}\email{dragos.i.palade@gmail.com}
    \affiliation{National Institute of Laser, Plasma and Radiation Physics,
	PO Box MG 36, RO-077125 M\u{a}gurele, Bucharest, Romania}
	\author{V.Baran}\email{baran@nipne.ro}
	\affiliation{Faculty of Physics, University of Bucharest, Romania}
	
	\keywords{Random Phase Approximation, Separable, Clusters}
	
	\begin{abstract}
		
		Starting from the Random Phase Approximation (RPA), we generalize the schematic model of separable interaction defining subspaces of $ph$ excitations with different coupling constants between them.
		This \emph{ansatz} simplifies the RPA eigenvalue problem to a finite, small dimensional system of equations which reduces the numerical effort. Associated dispersion relation and the normalization condition are derived for the new defined unknowns of the system. In contrast with the standard separable approach, the present formalism is able to describe more than one collective excitation even in the degenerate limit. The theoretical framework is applied to neutral and singly charged spherical sodium clusters and $C_{60}$ fullerene with results in good agreement with full RPA calculations and experimental data.
		
	\end{abstract}

	\maketitle
	
	\section{Introduction}
	\label{Intro}
	
	Recently \cite{PhysRevC.91.054303,baranromj} we have investigated the collectivity of the Giant Dipole Resonance and of the low-lying modes (Pigmy Dipole Resonance) in nuclear systems using the separable RPA. In contrast with the standard schematic model \cite{brown1959dipole} which employs separable residual interaction with a single specific coupling constant between nucleons, we have assumed the existence of two subsystems of particle-hole states and, consequently, three distinct coupling constants.
	
	In the present work we generalize this approach to a variable number of subsystems of particle-hole pairs, allowing us to describe an even higher number of collective resonances - a feature which is relevant to systems such as metallic clusters.
	
	The optical response of clusters has been intensively studied in the past three decades from both an experimental and a theoretical point of view (for reviews see \cite{kresin1992collective,deheer}). 
	
	The most pregnant effect in the optical response of large (metal) clusters, is the \emph{surface plasmon}. This represents a dipolar motion of the electron cloud against the ionic background, similar with the Goldhaber-Teller \cite{goldhaber1948nuclear} description of the Giant Dipole Resonance in nuclei. It is interpreted as a collective phenomena given by the superposition of single particle effects at nearly the same frequency. Another collective resonance is the so called \emph{volume plasmon} which, from a macroscopic view is characterized by density variations inside the cluster volume as in the Steinwedel-Jensen model \cite{steinwedel1950z}. 
	
	These are gross properties of the optical spectra, exhausting large fractions from the total sum rule in large or spherical shaped clusters. Still, spanning different sizes, the shell effect becomes stringent \cite{brack1993physics}, similar with the atomic case, allowing for the existence of many deformed clusters in between the spherical ones. Moreover, for clusters with a small number of atoms, the details of the ionic background become important. For all that reasons, we face in optical spectra relevant single particle excitations or fragmentations of the collective surface plasmon. The latter is connected with the phenomenon of Landau fragmentation.
	

	Unlike in nuclear physics, where Hartree-Fock (HF) is the main microscopic theory, the dynamics of electrons in clusters is usually investigated by means of the Density Functional Theory (DFT)
	\cite{van2001key,brack1993physics,calvayrac1997spectral} which is implemented through the Time-Dependent Kohn-Sham equations \cite{Kohnsham1965}. The extension to the time-dependent case describes the properties of excited states and its small amplitude limit represents the random phase approximation (RPA) correspondent of the HF derived RPA. 
	
	Even though RPA can be easier to implement than full HF or DFT, it still poses a difficult numerical task. This is reflected in the dimension of the space of single particle excitations, given the fact that for each pair of $ph$ states, two $6D$ integrals should be computed. The difficulty arises especially in deformed clusters, where the space of the single particle states lacks degeneracy and the spherical symmetric is removed.
	
	To overcome such challenges, there has been an active research in the past decades in the art of simplified RPA methods. The lowest level of approximation is given in the frame of sum rule techniques \cite{lipparini1991collective} where a single resonance is described. A more pertinent description is achieved with the local RPA method \cite{reinhard1990random} where a set of local operators is used to describe an equal set of collective resonances, but the Landau damping features are lost. The original separable ansatz \cite{brown1959dipole} is used  in cluster physics as the \emph{vibrating potential model} \cite{lipparini1991collective,nesterenko1995collectiveelambda}. This has the main disadvantage that only one collective excitation is obtained and the natural fragmentation of the surface plasmon is not properly described.
	
	A more refined method is provided by the \emph{self-consistent separable RPA} (SRPA) \cite{nesterenko1997multipole, nest2002} which combines the idea of more local operators and the separability ansatz of residual interaction. 
	
	In this work we generalize the separability of the residual interaction in the frame of DFT-LDA using a single operator but many coupling constants prescribed by $N$ blocks of $ph$ excitations, method which will be referred to as $N$ Block Separable RPA ($N$-BSRPA). Our formalism has many similar points with the one developed in \cite{nest2002}, but, as we shall see, the interpretations are fundamentally distinct. While SRPA describes a system of coupled collective motions, $N$-BSRPA accounts for a collection of coupled subsystems of oscillators.
	
	The paper is organized as follows: In Section \ref{theory} a short derivation of DFT derived RPA is given and the formalism of $N$-BSRPA is developed. In Section \ref{results} the analytical results are applied for some sodium clusters and $C_{60}$ fullerene. 
	
	\section{Theory}
	\label{theory}
	
	\subsection{RPA from DFT}
	\label{rpafromdft}

	The Time-Dependent DFT \cite{runge1984time} within the adiabatic local density approximation (LDA) \cite{bauernschmitt1996treatment} can be represented through the Kohn-Sham equations, which for a system of electrons described by single particle orbitals $\{\phi_j(\mathbf{r},t)\}$, read:
	
	\begin{align}\label{KSeq}
	h[\rho(\mathbf{r},t)]\phi_j(\mathbf{r},t) =i\hbar\partial_t\phi_j(\mathbf{r},t),
	\end{align}
	with 
	\begin{equation}
	\rho(\mathbf{r},t)=\sum_{j\in occ}|\phi_j(\mathbf{r},t)|^2.
	\end{equation}
	$j\in occ$ indexes the occupied single particle states, $\rho(\mathbf{r},t)$ is the total density of the particles and $h[\rho(\mathbf{r},t)]$ is a mean field Hamiltonian incorporating, besides the kinetic term, the Coulomb interaction between electrons, a local potential for the exchange-correlation effects and an external potential. Since we are interested in the normal modes of the electron system, we consider that the external potential is being given only by the Coulomb interaction between electrons and ions. 
	
	In the stationary case, the KS equations become eigenvalue problems $h[\rho_0(\mathbf{r})]\varphi_j(\mathbf{r}) =\varepsilon_j\varphi_j(\mathbf{r})$. Each orbital can be expanded around its stationary value $\phi_j(\mathbf{r},t) =(\varphi_j(\mathbf{r})+\delta
	\phi_j(\mathbf{r},t))e^{-i\varepsilon_jt/\hbar}$   and with this expansion we can linearize the Hamiltonian and the density as:
	
	\begin{equation}
	\label{linear}
	\begin{aligned}
	\delta\rho(\mathbf{r},t) &{}=2\sum_{j\in occ}Re[\varphi_j(\mathbf{r}) \delta\phi_j^*(\mathbf{r},t)],\\
	h[\rho] &{} =h_0[\rho_0]+\left.\frac{\delta h}{\delta\rho}\right\vert_{\rho_0}\delta\rho.\\
	\end{aligned}
	\end{equation}
	
	Since we are interested in the oscillating behaviour of $\delta\phi_j$ we search for the harmonic coefficients of the expansion in the basis of both occupied and unoccupied single-particle states:
	
	\begin{eqnarray}\label{expansion}
	\delta\phi_j(\mathbf{r},t)=\sum_{\mu\in all}\varphi_{\mu}(\mathbf{r})\left(X_{j\mu}e^{-iEt/\hbar}+Y_{j\mu}^*e^{iEt/\hbar}\right).
	\end{eqnarray}
	
	Replacing the linearised forms \ref{linear} and the expansion \ref{expansion} of $\delta\phi_j(\mathbf{r},t)$ in eq. \ref{KSeq} and integrating the equation after multiplication with $\varphi_{\nu}$, we obtain the usual form of the RPA eigenvalue equations 
	
	\begin{equation}\label{RPAeq}
	\begin{aligned}
	\varepsilon_iX_i^{(k)} +\sum_j(A_{ij}X_j^{(k)}+B_{ij}Y_j^{(k)}) &=E^{(k)}X_i^{(k)},\\
	\varepsilon_iY_i^{(k)} +\sum_j(B_{ij}^*X_j^{(k)}+A_{ij}^*Y_j^{(k)})&=-E^{(k)}Y_i^{(k)},
	\end{aligned}
	\end{equation}
	with $\sum_j(|X_{j}^{(k)}|^2-|Y_{j}^{(k)}|^2)=1$ as normalization condition .
	In deriving the above equations, we used the compact notations $X_{j\mu}\equiv X_i$, $Y_{j\mu}\equiv Y_i$,  $\varepsilon_\mu-\varepsilon_{j} \equiv\varepsilon_i$, $A_{ij}=\langle\varphi_{\nu}\varphi_{j}|\delta h/\delta\rho|\varphi_{i}\varphi_{\mu}\rangle$ and $B_{ij}=\langle\varphi_{\nu}\varphi_{\mu}|\delta h/\delta\rho|\varphi_{i}\varphi_{j}\rangle$. The superscript $(k)$ labels the solutions of the eigenvalue problem. For legibility we shall suppress this notation whenever possible.
	In literature, the results of this derivation are also known as the \emph{linearized TDDFT}. More detailed discussions and related derivations can be found in \cite{muta2002solving,yabana2006real,reinhard1992rpa,reinhard1996sum}.
	
	It might seem that, unlike in the HF derived RPA, the system of eqs.\ref{RPAeq} describes transitions from occupied states into other occupied states, fact which violates the Pauli principle, but in practice, since these terms cancel each other, we will set them to $0$.
	While in the standard RPA the residual two-body interaction is written as a difference between direct and exchange terms, in the DFT-LDA derived RPA it is given by the kernel $\delta h(\mathbf{r})/\delta\rho(\mathbf{r}')$. 
	
	The system of equations \ref{RPAeq} can be written in matrix notation as:
	
	\begin{eqnarray}
	\label{eigen}
	\left( \begin{array}{cc}
	\mathbf{\hat{A}} & \mathbf{\hat{B}} \\
	-\mathbf{\hat{B}}^* & -\mathbf{\hat{A}}^* \end{array}\right)\left( \begin{array}{c}
	\mathbf{X} \\
	\mathbf{Y}  \end{array}\right)=\left( \begin{array}{cc}
	\hat{\mathcal{E}}_-(E) & 0 \\
	0 & \hat{\mathcal{E}}_+(E) \end{array}\right)\left( \begin{array}{c}
	\mathbf{X} \\
	\mathbf{Y}  \end{array}\right)
	\end{eqnarray}	
	where we defined the RPA matrices $\hat{\mathbf{A}}=\{A_{ij}\}$, $\hat{\mathbf{B}}=\{B_{ij}\}$, $\hat{\mathcal{E}}_{\pm}(E)=\{\delta_{ij}(E\pm\varepsilon_i)\}$ and the vectors $\mathbf{X}=\{X_i\}$, $\mathbf{Y}=\{Y_i\}$. In the next subsection we will employ the ansatz of separability performing a transformation on this system of equations.

	\subsection{Block-separable anzatz}
	\label{relaxed}
	
	The standard separable ansatz states that for all $ph,p'h'$ pairs, the two-body matrix elements can be factorized in terms of a one-body transition operator $\hat{Q}$: $A_{i,j}=\lambda Q^*_{i}Q_{j}$ where $Q_{i}=\langle \varphi_p|\hat{Q}|\varphi_h\rangle$. Already in \cite{PhysRevC.91.054303} we have relaxed this condition using three different coupling constants. We generalize this result to $N$ blocks of $ph$ excitations. In order to achieve such a description in an elegant manner, in the following we will adopt the matrix notation with compact indexes $i=ph$ and $j=p'h'$. 
	
	Let us consider the unique set of coupling constants $\lambda_{i,j}=A_{i,j}/(Q^*_{i}Q_{j})=B_{i,j}/(Q_{i}Q_{j})$. We construct the set of matrices and vectors: 
	
	\begin{align*}
	&\hat{\mathbf{\Lambda}}=\{\lambda_{ij}\},\\
	&\hat{\mathbf{Q}}=\{Q_i\delta_{ij}\},\\
	&\mathbf{Z}=\left( \begin{array}{rc}
	\hat{\mathbf{Q}}^* & \hat{\mathbf{Q}} \end{array}\right)
	\left( \begin{array}{c}
	\mathbf{X} \\
	\mathbf{Y}  \end{array}\right),\\
	&\hat{\chi}(E)=|\hat{\mathbf{Q}}|^2[\hat{\mathcal{E}}_-^{-1}(E)-\hat{\mathcal{E}}_+^{-1}(E)],
	\end{align*}
	
	where $\hat{\mathbf{\Lambda}}$ is the real and symmetric matrix of coupling constants and $\hat{\mathbf{Q}}$ the diagonal matrix of $ph$ strengths. The $\mathbf{Z}$ vector hides more physical meaning being related to the transition amplitude for the $k-th$ excited state $\langle 0|\hat{Q}|k\rangle=|\mathbf{Z}^{(k)}|=\sum_i Z_i^{(k)}$. The element of the diagonal matrix $\hat{\chi}_i(E)=2|Q_i|^2\varepsilon_i(E^2-\varepsilon_i^2)$ can be identified as the linear response function of the system for a single $ph$ state.
	
	Using the separable form of the interaction matrices $\hat{\mathbf{A}}=\hat{\mathbf{Q}}^*\hat{\mathbf{\Lambda}}\hat{\mathbf{Q}}$ and $\hat{\mathbf{B}}=\hat{\mathbf{Q}}\hat{\mathbf{\Lambda}}\hat{\mathbf{Q}}$ we can rewrite the system of equations \ref{eigen} as:
	
	\begin{eqnarray}\nonumber
	\label{eigen2}
	\left( \begin{array}{cc}
	\hat{\mathcal{E}}_-^{-1} & 0 \\
	0 & \hat{\mathcal{E}}_+^{-1} \end{array}\right)
	\left( \begin{array}{cc}
	\hat{\mathbf{Q}}^*\hat{\mathbf{\Lambda}}\hat{\mathbf{Q}} & \hat{\mathbf{Q}}\hat{\mathbf{\Lambda}}\hat{\mathbf{Q}} \\
	-\hat{\mathbf{Q}}^*\hat{\mathbf{\Lambda}}\hat{\mathbf{Q}}^* & -\hat{\mathbf{Q}}\hat{\mathbf{\Lambda}}\hat{\mathbf{Q}}^* \end{array}\right)\left( \begin{array}{c}
	\mathbf{X} \\
	\mathbf{Y}  \end{array}\right)=\left( \begin{array}{c}
	\mathbf{X} \\
	\mathbf{Y}  \end{array}\right).
	\end{eqnarray}
	
	Multiplying both sides with $(\hat{\mathbf{Q}}^*,\hat{\mathbf{Q}})$ we obtain the simplified form:
	
	\begin{eqnarray}
	\label{eigen3}
	\hat{\chi}(E)\hat{\mathbf{\Lambda}}\mathbf{Z}=\mathbf{Z}. \end{eqnarray}
	
	Note that this relation is equivalent with the original RPA equations \ref{RPAeq} and since we defined the coupling constants for each $ph,p'h'$ pair, no approximation has been done yet. Even though we halved the dimensionality of the problem, the new dispersion relation is more complicated, being contained in the condition of singular matrix:
	
	\begin{eqnarray}
	\label{dispersion1}
	\det\left(\hat{\mathbf{I}}-\hat{\chi}(E)\hat{\mathbf{\Lambda}}\right)=0.
	\end{eqnarray}
	
	Let us define the set of single particle excitations as $\pi=\{|ph\rangle\}$ and decompose it in $N$ blocks $\pi=\bigcup_{i=1}^N\pi_i$.  
	We now employ the generalized concept of separability.
	We assume that pairs of particle-hole states interact with different coupling constant corresponding to the blocks they belong to $\lambda_{ij}\equiv\lambda_{nm},\forall i\in\pi_n, j\in\pi_m$, $\forall n, m=1,N$. With this \emph{ansatz} one can rearrange the elements from \ref{eigen3} such that the $\hat{\mathbf{\Lambda}}$ matrix becomes an array of constant blocks:
	
	\begin{eqnarray}
	\label{block}
	\hat{\mathbf{\Lambda}}=\left( \begin{array}{cccc}
	\lambda_{11}\mathbf{\hat{1}}_{11} & \lambda_{12}\mathbf{\hat{1}}_{12} & \cdots & \lambda_{1N}\mathbf{\hat{1}}_{1N} \\
	\lambda_{21}\mathbf{\hat{1}}_{21} & \lambda_{22}\mathbf{\hat{1}}_{22} & \cdots & \lambda_{2N}\mathbf{\hat{1}}_{2N} \\
	\vdots & \cdots & \cdots & \vdots \\
	\lambda_{N1}\mathbf{\hat{1}}_{N1} & ... & ... & \lambda_{1N}\mathbf{\hat{1}}_{NN} \\
	\end{array}\right),
	\end{eqnarray}
	
	where $\mathbf{\hat{1}}_{nm}$ is a matrix with all elements $1$ and dimension $|\pi_n|\times |\pi_m|$, where $|\pi_i|$ is the cardinality of set $\pi_i$. Thus, the system \ref{eigen3} is reduced to an equivalent $N-$dimensional system (with the same form),
	in which the quantities involved are now defined as sums over $ph$ excitations in the same block $\chi_n(E)=\sum_{j\in\pi_n}\chi_j(E)$, $Z_n=\sum_{j\in\pi_n}Z_j$, $|Q|^2_{nm}=\delta_{nm}\sum_{j\in\pi_n} |Q|_j^2$, $n=\overline{1,N}$. 
	
	The same system of equations can be obtained using the second quantization formalism and imposing the following separability of the residual interaction:
	
	\begin{align}
	\label{secondq}
	&\frac{\delta\hat{h}}{\delta\rho}=\frac{1}{2}\sum_{n,m}\lambda_{nm}\hat{z}^\dagger_n \hat{z}_m,\\
	&\hat{z}_n=\sum_{j\in\pi_n}Q_j \hat{c}_j,
	\end{align}
	
	where $\hat{c}^\dagger_j\equiv\hat{a}^ \dagger_p\hat{a}_h$. It can be shown that the vector $Z$ is related to the matrix elements of the operator $\hat{z}_n$ by $Z_n^{(k)}=\langle 0|\hat{z}_n|k\rangle$, and that the operators $\hat{z}_n$ verify the anti-commutation relation  $\{\hat{z}^\dagger_n,\hat{z}_m\}=\delta_{nm}(\sum_{i\in\pi_n}|Q_i|^2)=|\hat{\mathbf{Q}}|^2_{nm}$ 
	Using the normalization of the initial RPA $\sum_j(|X_{j}^{(k)}|^2-|Y_{j}^{(k)}|^2)=1$ and the dependency between $\mathbf{Z}$ and $\{\mathbf{X,Y}\}$ we can derive a normalization condition for the solutions of the \ref{eigen2}:
	
	\begin{eqnarray}
	\mathbf{Z}^\dagger\partial_E\hat{\chi}^{-1}(E)\mathbf{Z}=-1.
	\end{eqnarray}
	
	Recognizing that the elements of $\hat{\chi}$ are response functions for each block $\pi_n$ and having in mind the form of the residual interaction \ref{secondq} we support the idea that the present formalism models a system of coupled subsystems of oscillators.
	
	Let us now stress out the main advantages of $N-$BSRPA:
	
	\begin{enumerate}
		\item The dimension of the original problem $2occ\times unocc$ (which is infinite in principle) can be dramatically reduced to a system with dimension $N$.
		\item The number of solutions of the dispersion relation \ref{dispersion1} is preserved since the latter can be written as an $2occ\times unocc$ real polynomial in $E$. The rank-nullity prescribes one $Z^{(k)}$ solution for each $E^{(k)}$ excitation energy.
		\item The phenomenon of Landau damping is reproduced due to the presence of single particle effects in the block response function $\chi_n(E)$.
		\item The TKR \cite{TKR} sum rule is recovered by the solutions of \ref{eigen2}:$$m_1(Q)=\sum_i\left|Q_i\right|^2\varepsilon_i=\sum_k\left|Z^{(k)}\right|^2E^{(k)}.$$
	\end{enumerate}

	The goal of the whole RPA procedure is to reproduce the linear response function of the system. Since we prescribed $\hat{Q}$ as being the transition operator, the total response function (normalized to the sum rule) can be written as:
	
	\begin{eqnarray}
	\label{spectrum}
	\mathcal{S}(E)=\sum_k\frac{\left|Z^{(k)}\right|^2}{m_1(Q)}\mathcal{L}\left(E,E^{(k)},\Gamma^{(k)}\right),
	\end{eqnarray} 
	
	where $\mathcal{L}\left(E,E^{(k)},\Gamma^{(k)}\right)=2E^{(k)}[(E+i\Gamma^{(k)})^2-E^{(k)2}]^{-1}$ is the Lorentzian broadening of the delta function with some phenomenological width $\Gamma^{(k)}\ll E^{(k)}$. While this is the standard procedure in the representation of optical spectra (for example) we should be aware that the block approximation involves basically an averaging over coupling constants, therefore, the broadening can be explained as a feature of the dispersion of real $\lambda_{ph,p'h'}$.

	We stress out that even if our results are very similar to the ones in \cite{PhysRevC.91.054303} (Section II D), the physical picture is completely different: while they use a set of operators projected on the $\pi$ space of $ph$ excitations, we use a single operator projected on a set of $\pi_n$ spaces. Consequently, the interpretation of the coupling constants is different. The similarities arise from the mathematical apparatus which is basically the same.
	
	\subsection{Particular cases}
	\label{specific}
	
	Let us observe that the original separable RPA \cite{brown1959dipole} is recovered if $N=1$, i.e. one single block of relevant excitations is considered, which implies an unique coupling constant $\lambda$ between any pair of $ph$ excitations. Then the dispersion relation \ref{dispersion1} reduces to the well known equation: 
	
	\begin{eqnarray}
	\frac{1}{\lambda}=\sum_i\frac{2|Q_i|^2\varepsilon_i}{E^2-\varepsilon_i^2}.\end{eqnarray}
	
	The $N=2$ case has been treated in our previous work \cite{PhysRevC.91.054303}. At the other limit, when $N=occ\times unocc$, i.e. each block contains a single $ph$ excitation, the standard RPA is recovered.
	
	Also, the Tamm-Dancoff approximation can be obtained in our formalism by setting $\mathbf{Y}=0$ and defining the response function matrix as $\hat{\chi}_n=\sum_{i\in\pi_n}|Q_i|^2/(E-\varepsilon_i)$. 
	
	We note that in the limit $E\gg\varepsilon_i$ the determinant from the dispersion relation \ref{dispersion1} can be expanded in the first order and the equation becomes:
	
	\begin{eqnarray}
	\label{dispersion2}
	1-Tr\{\hat{\chi}(E)\hat{\mathbf{\Lambda}}\}=1-\sum_n\lambda_{nn}\chi_n(E)=0.
	\end{eqnarray}
	
	This means that for high energies, the interaction between blocks of $ph$ states can be neglected, in other words, the high energy excitations can be viewed schematically as collection of non-interacting subsystems (oscillators). These case can be justified keeping in mind that in clusters, the relevant single particle energies (those with large strength $|Q_{ph}|$) $\varepsilon_{ph}\simeq 1eV$ while the collective states for the surface/volume plasmons are characterized by $E_{s/v}\simeq 3eV/4eV$. 
	
	In nuclear physics it is customary to consider the limit of a completely degenerate spectrum. This is motivated by the fact that the isotropic harmonic oscillator, which has equally distanced occupied levels of energy, represents an empirically verified description of the mean field potential of the system in the ground state. Consequently, the relevant $ph$ excitations are between equidistant shells. The same approximation is also valid in cluster physics. In this limit, $\varepsilon_i\equiv\varepsilon, \forall i\in \pi$, the system of equations \ref{eigen3} becomes:
	
	\begin{eqnarray}
	\label{degeneratule}
	|\hat{\mathbf{Q}}|^2\hat{\mathbf{\Lambda}}\mathbf{Z}=\frac{E^2-\varepsilon^2}{2\varepsilon}\mathbf{Z}.
	\end{eqnarray}
	
	It follows that the solutions of the dispersion relation can be written as $E^{(k)}=\pm\sqrt{\varepsilon^2+2\varepsilon q_k}$, where $\{q_k\}$ are the eigenvalues of the $|\hat{\mathbf{Q}}|^2\hat{\mathbf{\Lambda}}$ matrix. Moreover, we obtain a simple normalization condition  $\mathbf{Z}_k^\dagger|\hat{\mathbf{Q}}|^2\mathbf{Z}_k=E^{(k)}/\varepsilon q_k^3$. Being obtained in the frame of degenerate limit, all $N$ solutions describe collective excitations.

	At this end, we recognize that in our previous work we were not able to identify the connection between the analytical solutions derived in there and the fact that they were eigenvalues of a certain matrix ($\mathbf{|\hat{Q}|^2\hat{\Lambda}}$).
	
	Finally, we should link the generalized static polarizability for the $\hat{Q}$ operator with solutions provided by our formalism :
	
	\begin{eqnarray}
	\label{polar}
	\alpha(Q)=\sum_k^N\frac{|Z^{(k)}|^2}{E_k}
	\end{eqnarray} 
	since it will serve as a supplementary test of the method. 

\section{Results and discussion}
\label{results}

Before presenting the numerical results, we will briefly describe the method used for the calculation of the ground state electronic spectrum. 

It is known \cite{brack1993physics} that the response of clusters to external perturbations in the linear regimes is dictated by the valence electrons, while the core electrons and the nuclei remain frozen. The ionic background can be modelled as a positive charge distribution

\begin{align}
\rho_{jel}(\mathbf{r})=\rho_0[1+\exp(\sigma(|r|-R_{jel}))]^{-1},
\end{align}
with
\begin{equation}
R_{jel}=R_0\left(1+\sum_{lm}\alpha_{lm}Y_{lm}(\Omega)\right).
\end{equation}
$\rho_0$ is equal to the ratio $3n/4\pi r_s^3$, where $r_s$ is the Wigner-Seitzs radius. In practice, this value is adjusted such that $\rho_{jel}$ obeys the normalization condition $\int d^3\mathbf{r}\rho_{jel}(\mathbf{r})=n$, where $n$ is the number of atoms in cluster. The parameters $R_0=r_sn^{1/3}$ and $\alpha_l^{m*}=\alpha_l^{-m}$ are a measure of the deformation of the cluster and $\sigma$ replaces the diffusive character of the jellium which is known to give better results than the step profile ($\sigma\to\infty$). 

The ground state of the system was obtained by solving the DFT-LDA equations

\begin{eqnarray}\label{KS}
&\left(-\frac{\hbar^2}{2m}\nabla^2+V_{KS}(\mathbf{r})\right)\phi_j(\mathbf{r})= \varepsilon_j\phi_j(\mathbf{r}),\\
&V_{KS}(\mathbf{r})=V_{ion}(\mathbf{r})+V_H(\mathbf{r})+V_{xc}(\mathbf{r}).
\end{eqnarray} 

$V_{ion}$ is the potential created by the ionic background (either from pseudopotentials, or the jellium model), the Hartree potential $V_H$ is subject to Poisson equation $\nabla^2 V_H=-4\pi\rho(\mathbf{r})$ and the exchange-correlation potential $V_{xc}$ is approximated with the Gunnarson-Lundqviust functional \cite{gunarsson}:

\begin{eqnarray}
V_{xc}[\rho]=-\left(\frac{3\rho}{\pi}\right)^{1/3}-0.0333 \ln\left(1+11.4\left(\frac{4\pi\rho}{3}\right)^{1/3}\right).
\end{eqnarray}

Numerically, we solved the eigenvalue problems \ref{KS} for general clusters using a code that discretizes the equations on a $3D$ uniform grid with a step of $\simeq 0.1R_0$ for a box with a length 5 times the radius $R_0$. The eigenproblem is solved using the imaginary time method coupled with the split-operator technique. For the kinetic operator, a Fourier representation has been employed, as well as for the computation of the Coulomb potential. For simplicity, we consider the jellium representation, ignoring the ionic configuration.
For more computational details see \cite{Calvayrac2000493,YABANAREAL}.

The residual two-body interaction prescribed by \ref{KS}  can be written explicitly as:

\begin{eqnarray}
\frac{\delta h}	{\delta \rho}=\frac{e^2}{|\mathbf{r-r'}|}+\frac{\delta v_{xc}(\mathbf{r})}{\delta\rho(\mathbf{r'})}\delta(\mathbf{r-r'})
\end{eqnarray}

Now, the numerical difficulty posed by RPA is obvious, being reflected in the $A_{ph,p'h'}$ terms which are computed as $6D$ integrals for the Coulomb kernel. The LDA nature of the exchange-correlation potential makes the evaluation of matrix elements simpler as $3D$ integrals. In principle, the long-range character of the Coulomb interaction makes the assumption of separability impossible to prove. The level of applicability should be described only by numerical simulations of the true RPA matrix elements and comparison with their separable form. In practice the coupling constants $\lambda_{ph,p'h'}$ are computed for each $ph-p'h'$ pair.

\subsection{Sodium clusters}
\label{sodium}

The sodium clusters are known to be textbook cases in cluster physics due to their alkali character and for that we start with them as test cases. The value of $r_s\approx 3.93 a.u.$ is used in the calculation of $R_0$. The spherical shaped clusters $Na_{n(+1)}^{(+)}$ (with a \emph{magic number}: $n=2,8,20,40,58,92,138,..$ of atoms) are easier to tackle numerically, both from the perspective of DFT-LDA as well as of RPA  calculations (for details, see \cite{yannouleas1993evolution}). Also, their optical spectra exhibits a strong surface plasmon around $3 eV$

\begin{figure}
	\centering
	\includegraphics[width=1.0\linewidth]{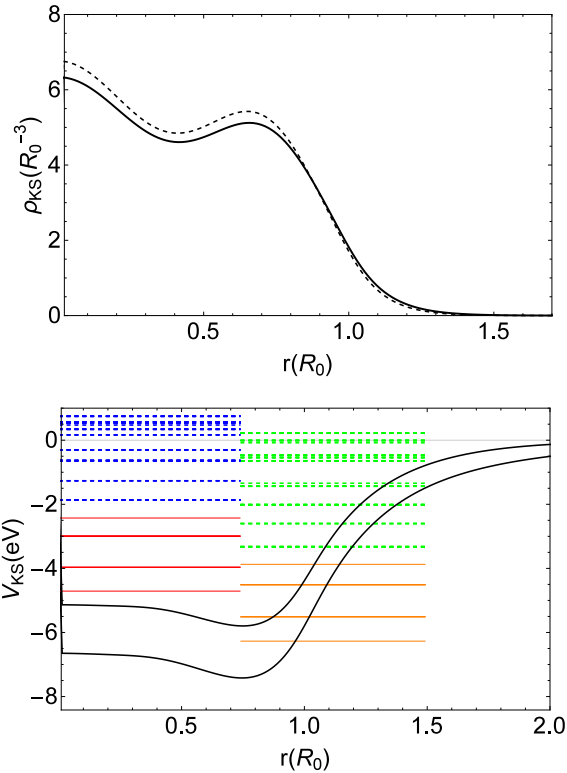}
	\caption{The electronic radial density, KS potential and first energetic eigenvalues are plotted for $Na_{20}$ and for singly charged $Na_{21}^+$ as it follows: in the upper figure the density solid line-$Na_{21}^+$ and dashed line $Na_{20}$; in the bottom figure the occupied states (red-$Na_{20}$, orange-$Na_{21}^+$) and the first $120$ unoccupied states (blue-$Na_{20}$, green-$Na_{21}^+$)}
	\label{fig:1}
\end{figure}

The numerical challenge comes in deformed clusters where the natural degeneracies from the spherical case are removed and the spectrum of needed $ph$ excitations considerably enlarged. This increases the dimension of the RPA system of equations and consequently the number of matrix elements to be computed. In turn, as in nuclear physics, the deformation induces a splitting of the strong surface plasmon (landau fragmentation)

As a demonstration of the numerical results regarding the ground state we plot in Fig. \ref{fig:1} the radial profile of the electronic density in $Na_{20}$ and $Na_{21}^+$ as well as the first few energetic levels.

\begin{figure}[h]
	\centering
	\includegraphics[width=1.\linewidth]{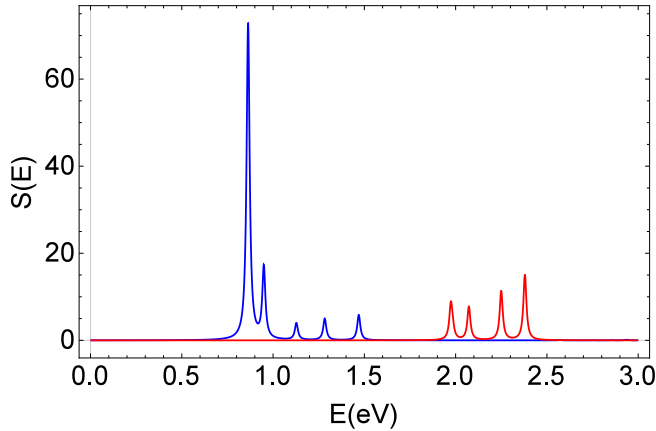}
	\caption{Single particle $ph$ excitation strengths in $Na_{58}$ for the dipole operator (blue) and volume operator (red))}
	\label{fig:3}
\end{figure}

The formalism described in the previous section is general and valid for any transition operator $\hat{Q}$ as long as the block-separable anzatz holds. In practice we should use the standard dipole operator to describe the optical spectrum, or, a more complex local operator of the following form: $Q(\mathbf{r})=\sum_{l,m}r^lY_{lm}(\Omega)$ \cite{Rowe_1970}. This operator accounts for dipolar surface oscillations and so, for the surface plasmon. To account for the volume plasmons we should employ operators which have non-zero laplacian character, generically $Q(\mathbf{r})=r^pY_{lm}(\Omega), p\neq l$. In the end we use the operators  $\hat{Q}_s=z$ for dipole normal modes and $\hat{Q}_v=r^2$ for volume modes.!!!

Information about relevant excitations in the system can be obtained by computing the spectrum $\mathcal{S}(E)$ for the independent particle response. In
the limit of $\hat{\mathbf{\Lambda}}\to\hat{0}$, the spectrum \ref{spectrum} reduces to the trace of $\hat{\chi}$. In Fig. \ref{fig:3} we have plotted this quantity for  $Na_{21}^+$ both for $Q_s$ and $Q_v$. 
The same behaviour can be seen in all sodium clusters. The degenerate limit is justified because of the narrow spectral region in which the matrix transition operator exhibits large values.
In particular we can compute the degenerate $ph$ energies as a weighted mean:  $\bar{\varepsilon}_{s/v}=\sum_{ph}\varepsilon_{ph}|Q_{ph}^{s/v}|^2/\sum_{ph}|Q_{ph}^{s/v}|^2$. 

\begin{figure}
	\centering
	\includegraphics[width=1.0\linewidth]{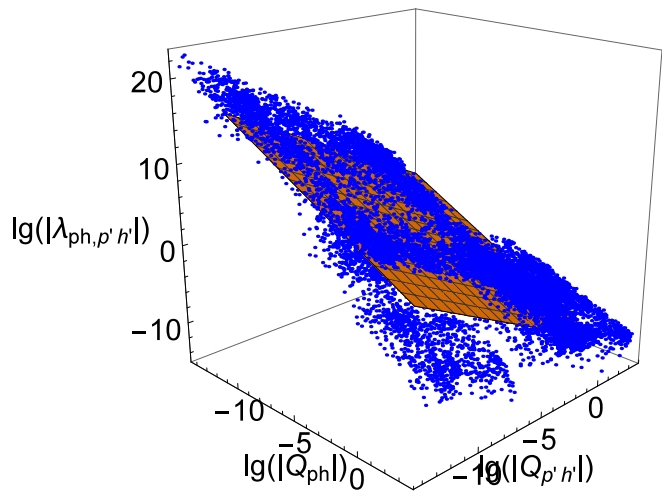}
	\caption{Scattered plot in $Na_9^+$ for $lg(|\lambda_{ph,p'h'}|)$ vs. $lg(|Q_{ph}|),lg(|Q_{p'h'}|)$ fitted with a plane}
	\label{fig:4}
\end{figure}

In order 
to prove the validity of the separable ansatz, we perform an analysis of the $\lambda_{ph,p'h'}$ set in respect with the characteristics of $ph$ states. Since the obtained values span many orders of magnitude we have plotted in Fig. \ref{fig:4} the logarithmic dependence between $\lambda_{ph,p'h'}$ and $|Q_{ph}|,|Q_{p'h'}|$ for $Na_9^+$. Surprisingly enough, we have found that the set of data points lies roughly in a plane. Moreover this type of behavior has been found in all the other simulated clusters. This has driven us to propose, without any mathematical justification, the following empirical relation $\lambda_{ph,p'h'}=c (Q_{ph}Q_{p'h'})^\gamma$, where $c>0$ and $\gamma<0$. 

\begin{figure}[h]
	\centering
	\includegraphics[width=1.0\linewidth]{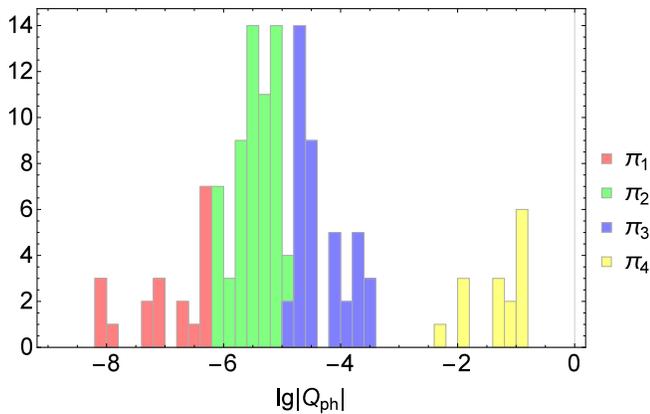}
	\caption{Distribution of $lg|Q_{ph}|$ in $Na_{9}^+$ for the dipole operator. The possible configuration of $4$ blocks $\pi_i$, $i=1,4$ is represented with different colours.}
	\label{fig:5}
\end{figure}

With these, it becomes obvious that the validity of the block ansatz relies on the possibility of defining the blocks from the distribution of $Q_{ph}$ elements. This is visually shown in \ref{fig:5} for $Na_{9}^+$ in a logarithmic histogram of $|Q_{ph}|$ where the a choice of $N=4$ blocks is represented in a colour plot.

As a conclusion of these features provided by the numerical computation of RPA $\lambda_{ph,p'h'}$ coupling strengths we propose the following scheme:

\begin{itemize}
	\item Solve the KS problem for the system and obtain the occupied and unoccupied spectrum.
	\item Compute the $\{\varepsilon_{ph}, Q_{ph}\}$ strengths for the single particle $ph$ excitations.
	\item Choose the $ph$ blocks $\pi_n$ separating regions in the distribution of $Q_{ph}$ strengths. 
	\item Using three free parameters $c, \gamma$ compute
	the block coupling constants $\lambda_{nm}=c (\langle Q\rangle_{n}\langle Q\rangle_{m})^\gamma$, where $\langle Q\rangle_{n}=\sum_{i\in\pi_n}Q_i/|\pi_n|$ is the mean in the block.
	\item Solve the dispersion relation \ref{dispersion1} and the system of eqns. \ref{eigen2} for block configuration
	\item Compute the polarizability accordingly with \ref{polar}
	\item Tune the free parameters such that the resulting polarizability $\alpha(Q)$ approaches the experimental value. 
	\item Compute the response function from \ref{spectrum}.
\end{itemize}

\begin{figure}
	\centering
	\includegraphics[width=1.0\linewidth]{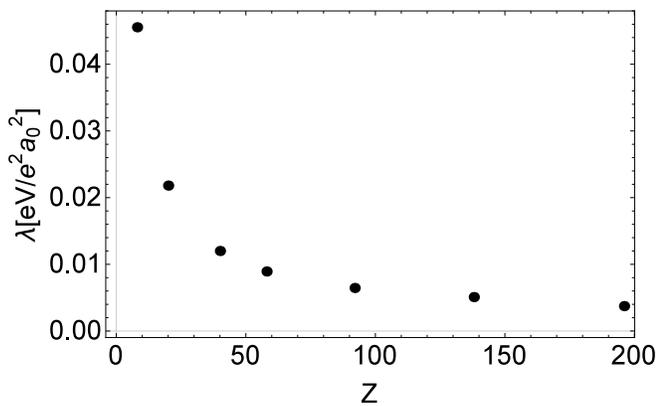}
	\caption{Values of $\lambda$ in the case of one-block and degenerate limit obtained from the DFT-LDA prescription for the degenerate energy and the experimental values for the peak of the surface plasmon \cite{XiaKresin2009}}
	\label{fig:6}
\end{figure}

In order to appreciate the correct magnitude of the coupling constants, we first apply the $N$-BSRPA in the degenerate limit of the case $N=1$. The degenerated energy is computed as above
$\bar{\varepsilon}=m_1(Q)/m_0(Q)$. 
We take $Q=e z$ the dipole operator and using the solutions for \ref{degeneratule} eigenvalue problem we compute the $\lambda$ that reproduce both the static polarizabilities (taken from \cite{tikhonovstatic}) and the centroid of the surface plasmons. The clusters $Na_n$ with $n=8,20,40,58,92,138,196$ have been chosen due to the fact that in deformed ones, the degeneracy is not a valid approximation anymore. The computation reads $\lambda=(E_{pl}^2-\bar{\varepsilon}^2)/(2\bar{\varepsilon}m_0)$ and the results can be seen in Fig. \ref{fig:6}. From a numerical fit, we have found the following values for the parameters of the empirical formula $\gamma\approx-0.75$ and $c\approx0.025 eV/(e^2a_0^2)$ (where $e$ is the electron's charge, $a_0$ the atomic unit of length and $Q_{ph}$ is expressed in $ea_0$).  

\begin{figure}
	\centering
	\includegraphics[width=1.0\linewidth]{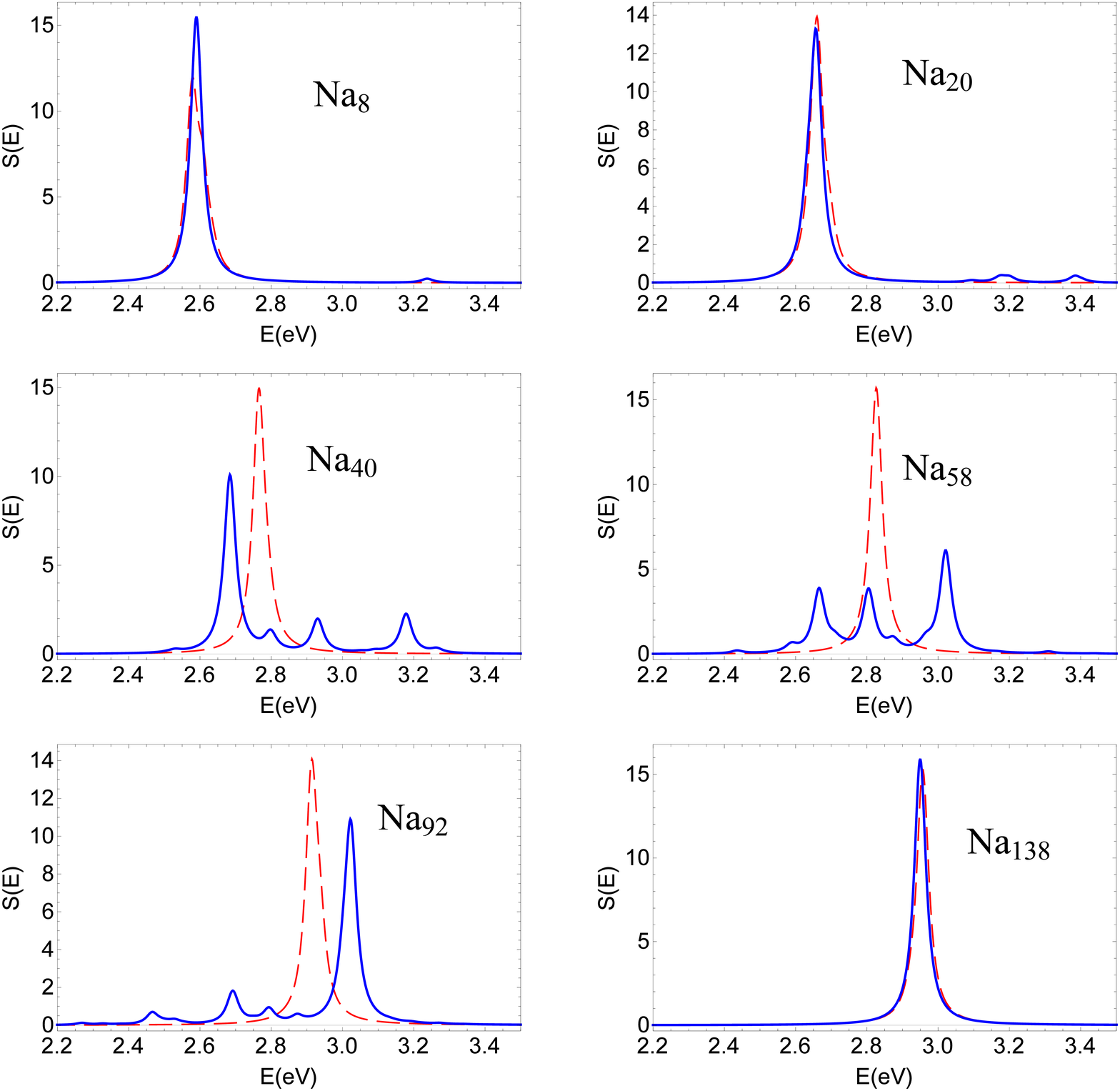}
	\caption{Optical spectra obtained with full RPA (red, dashed) and $1$-BSRPA (blue) in the one-block case for spherical clusters $Na_n$ where $n=8,20,40,58,92,138$}
	\label{fig:7}
\end{figure}

Going a step further, we disregard the degenerate limit and solve the $1$-BSRPA problem for the same neutral spherical clusters obtaining the optical spectra plot in Fig. \ref{fig:7} in respect with the true RPA results. A generic width $\Gamma_k=0.02eV$ has been used to broaden the resonances accordingly with eq. \ref{spectrum} as the lorentzian parameter. Also, we have used for $\lambda$ the values obtained in Fig. \ref{fig:6}. From the numerical results, a gross agreement can be found between full RPA and $N$-BSRPA even for this minimal case $N=1$. The differences can be explained recognizing that $N=1$ case is a crude approximation.

\begin{figure}
\centering
\includegraphics[width=1.0\linewidth]{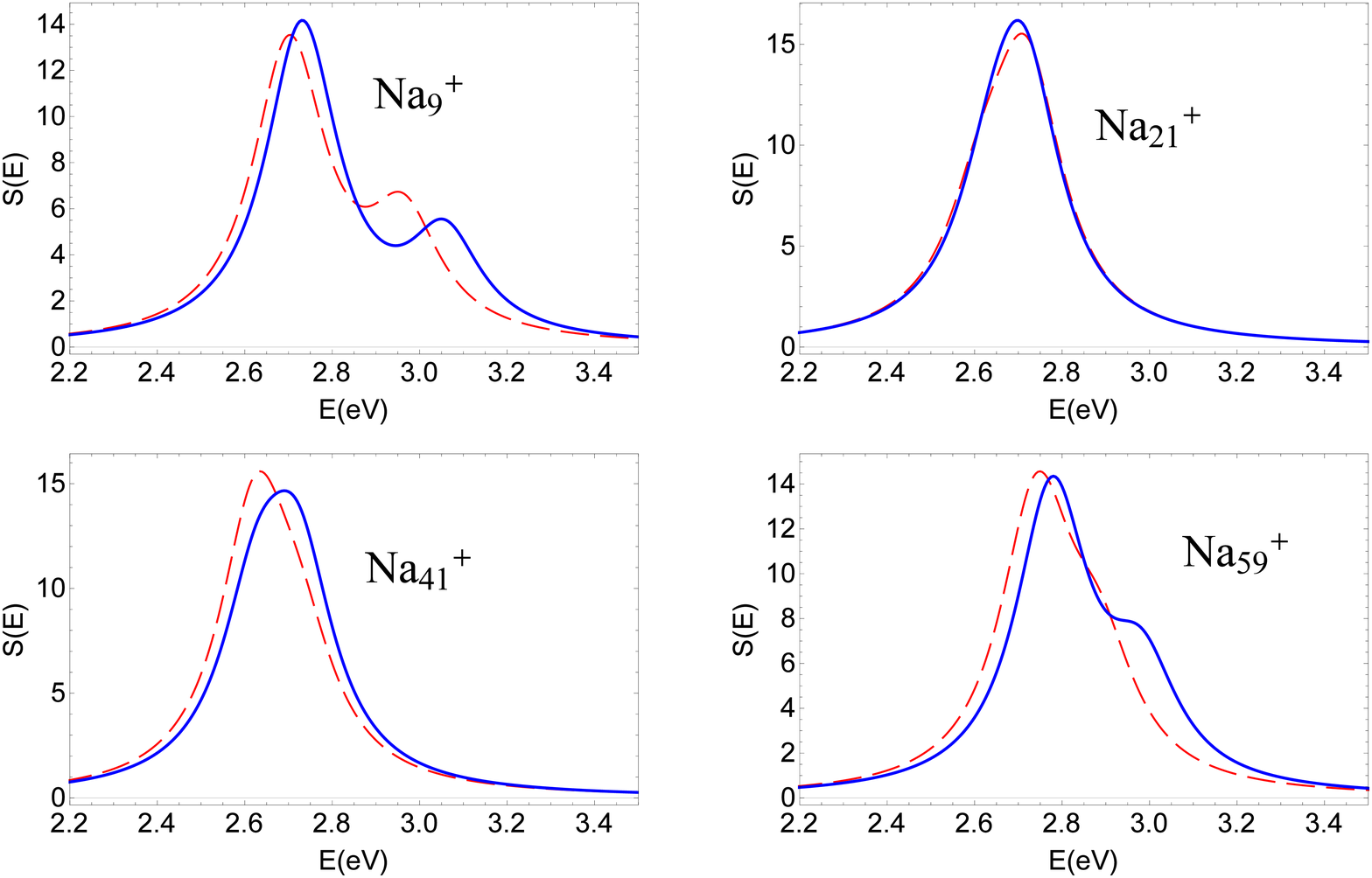}
\caption{Optical spectra obtained with full RPA (red, dashed) and with $4$-BSRPA in $Na_n^+$, where $n=9,21,41,59$}
\label{fig:8}
\end{figure}

In order to have a final test for $N$-BSPRA, we investigate the optical response for singly charged sodium clusters $Na_n^+$, $n=9,21,41,59$ which are also spherical symmetric. Even though, in principle, the most accurate results should be recovered for $N\to occ\times unocc$, we have found that the formalism with $N=4$ provides the best computational cost-accuracy ratio. For that reason we compute in Fig. \ref{fig:8} $\mathcal{S}(E)$ for the above mentioned clusters employing $N=4$ and using the discussed empirical parametrization for $\lambda_{nm}$. The parameter $c$ has been left free and tuned in order to obtain the RPA prescribed static polarizability as a first test. The results are compared with full RPA calculations which are in good agreement with the experimental data \cite{goodexp}. 
We conclude at this end that $N$-BSRPA gives a pertinent description of the dipole excitations in sodium clusters.

\subsection{$C_{60}$ Fullerene}
\label{c60}

Since the discovery of $C_{60}$ fullerene \cite{kroto1985c} a lot of experimental and theoretical attention has been drawn towards its spectacular properties. In particular, its optical spectrum is known to be dominated by two surface plasmons around $20 eV$ and $40eV$ \cite{scully2005}. In electron energy loss experiments three volumes plasmons emerge in the $20-30 eV$ range \cite{korolees}. The theoretical work done in this direction has been able to reproduce the experimental values from many perspectives as RPA \cite{bertschcollective1991}, ab-initio [], TDDFT \cite{scully2005}, Time-Dependent Thomas Fermi \cite{palade2015}, etc. 

Since it is genuinely a more complicated system than the simple alkali $Na$ clusters we considered it as a good further test case for the present formalism. 

\begin{figure}
	\centering
	\includegraphics[width=1.0\linewidth]{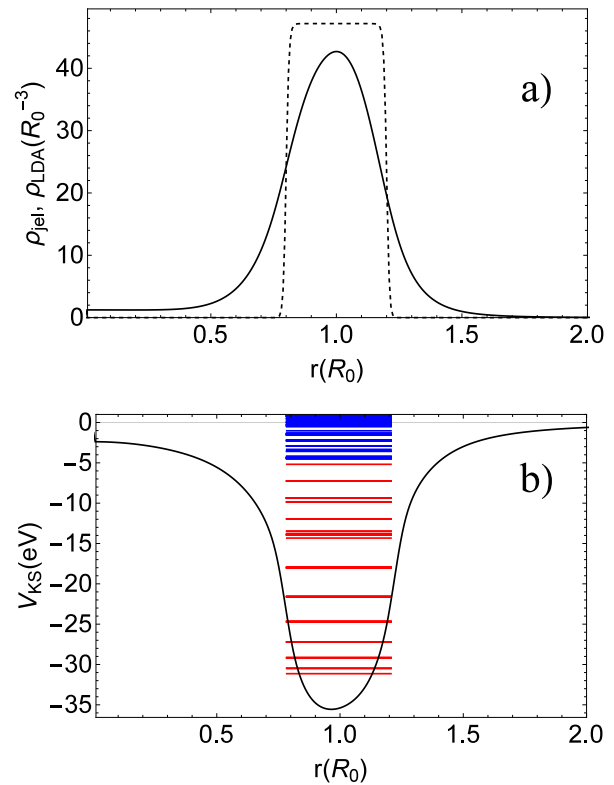}
	\caption{In figure a) is plotted the density of electrons (solid line) vs. the density of jellium (dashed line); Figure b) presents the KS potential (solid line), the occupied (red lines) states and first few unoccupied (blue lines) energetic levels}
	\label{fig:9}
\end{figure}

In order to solve the KS equations for the ground state we have modelled the ionic background accordingly with \cite{scully2005,madjet2008} defining a jellium shell $\rho_{jel}(\mathbf{r})\propto \Theta(r-r_1)\Theta(r_2-r)$ centred on the position of the carbon nuclei, where $r_1=2.8\mathring{A}$, $r_2=4.2\mathring{A}$ and normalized to $n=240$ which is the number of considered valence electrons. Following \cite{pusk1993photo} we have added a supplementary pseudopotential $V_0$ to the total KS potential in order to account for the ionic structure. Due to the spherical symmetry of the cluster of $C_{60}$ we were able to solve numerically only the radial KS problem and so, a more refined grid. For the calculation of matrix elements we have used the multipolar expansion of the Coulomb kernel (for more details about calculations of RPA in spherical systems see \cite{yannouleas1993evolution}). 

\begin{figure}[h]
	\centering
	\includegraphics[width=1.0\linewidth]{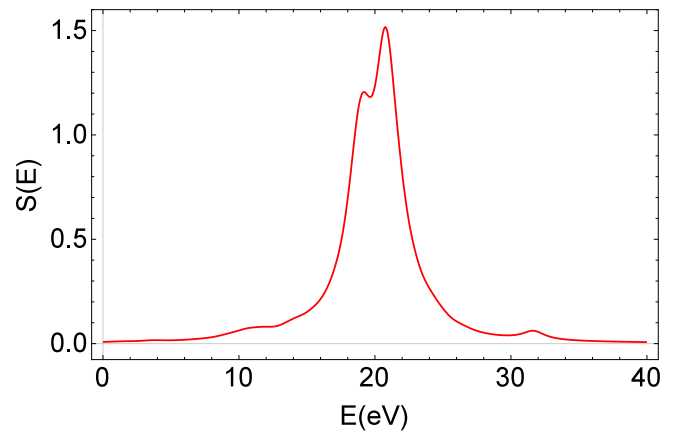}
	\caption{Dipole spectrum for $C_{60}$ obtained for $N=1$ block and $\lambda\approx 0.019$}
	\label{fig:10}
\end{figure}

A first set of results is plotted in Fig. \ref{fig:9} where the electronic density, the effective KS potential and the first few energetic levels are represented. 

Using the prescription for $\lambda$ obtained for sodium clusters we use the $1-BSRPA$ for $C_{60}$ to construct the optical spectrum in the case of dipole operator. The result can be seen in Fig. \ref{fig:C60zero} where the main surface plasmon from around $20eV$ is well reproduced. A spurious splitting is present as well as a second, smaller peak around $32eV$ which can be interpreted as the surface plasmon from around $38 eV$.

\begin{figure}
\centering
\includegraphics[width=1.0\linewidth]{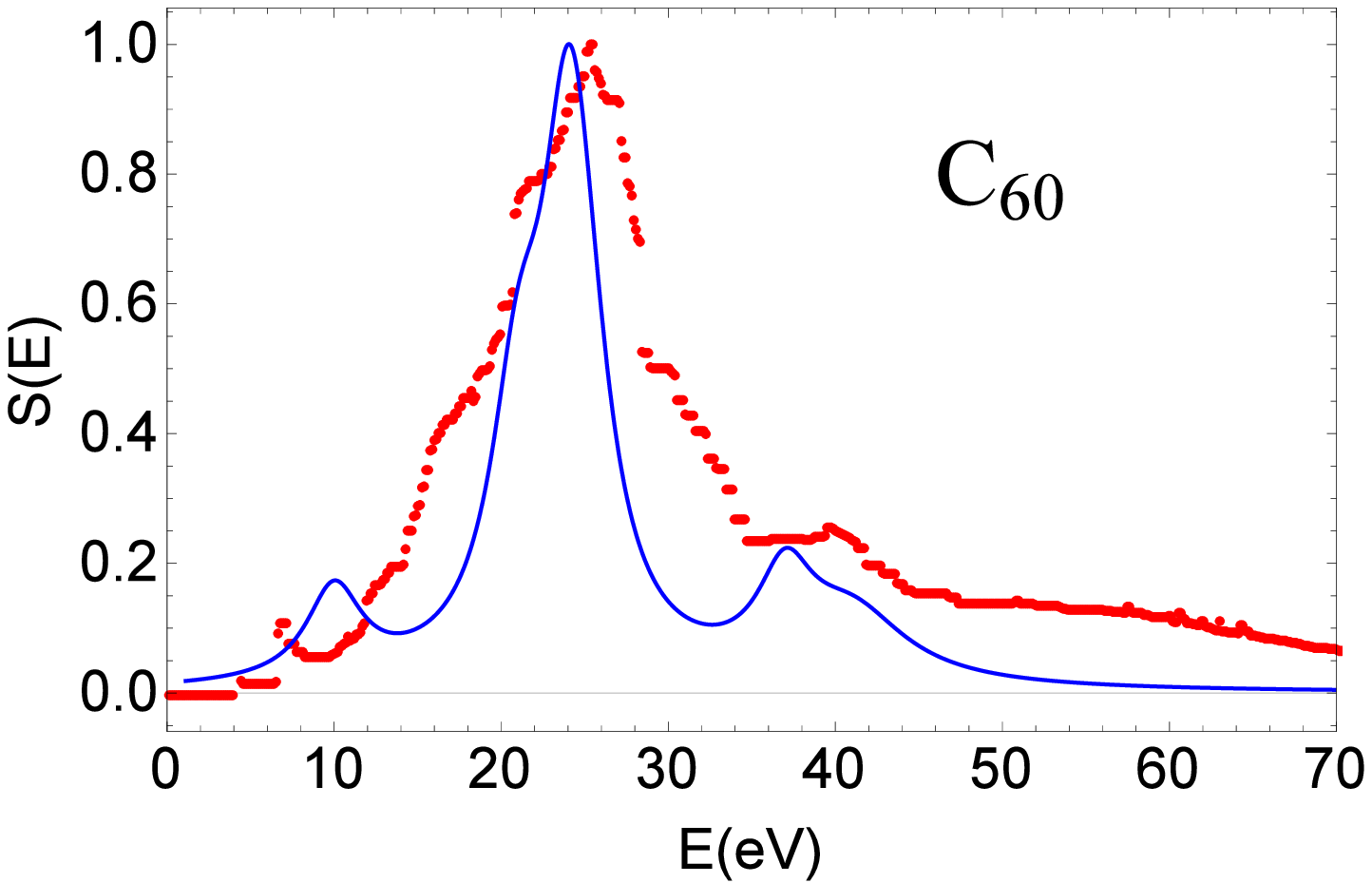}
\caption{}
\label{fig:11}
\end{figure}

Finally, we use the $5$-BSRPA with the $Q_s(\mathbf{r})=ez$ for the optical spectrum. The same parametrization has been used for $\lambda_{nm}$ as in the case of sodium clusters, and the free parameter $c$ has been tuned to reproduce the experimental static polarizability \cite{C60polar}. In Fig. \ref{fig:11} are reproduced the numerical results vs the experimental spectrum \cite{KafleC60}.

\section*{Conclusions}
\label{Conclusions}

Starting from DFT derived RPA equations we employ the ansatz of separable residual interaction defining $N$ blocks of $ph$ excitation with specific coupling constants between them. We show that, in the frame of this approximation, the RPA eigenvalue problem can be converted in a homogeneous system of $N$ equations. Associated dispersion equation and normalization condition is derived for the unknowns of the reformulated problem. 

The present $N$-BSRPA formalism has the main advantage over the standard RPA that reduces dramatically the dimensionality of the eigenvalue problem conserving the number of the solutions and consequently the Landau damping feature of the normal modes. Also, the nature of the collective excitations is naturally reflected in the degenerate limit of the system. On the other side, the formalism does not prescribe a recipe for the computation of coupling constants, which have to be tuned in order to reproduce the experimental data. Still, an empirical formula is proposed which holds in the case of sodium clusters and $C_{60}$ fullerene, but we do not sustain the idea that the parametrization is universal. 

We conclude that the present generalization of the separable RPA is a real alternative to the high costs of the full RPA calculations, but the problem of coupling constants remains open. We support the idea that the present formalism could find applications in many other systems from nuclear to solid state physics, but also in modelling other phenomena as the pairing is. 

\bibliographystyle{apsrev4-1}
\bibliography{bibRPA}

\begin{thebibliography}{40}%
\makeatletter
\providecommand \@ifxundefined [1]{%
 \@ifx{#1\undefined}
}%
\providecommand \@ifnum [1]{%
 \ifnum #1\expandafter \@firstoftwo
 \else \expandafter \@secondoftwo
 \fi
}%
\providecommand \@ifx [1]{%
 \ifx #1\expandafter \@firstoftwo
 \else \expandafter \@secondoftwo
 \fi
}%
\providecommand \natexlab [1]{#1}%
\providecommand \enquote  [1]{``#1''}%
\providecommand \bibnamefont  [1]{#1}%
\providecommand \bibfnamefont [1]{#1}%
\providecommand \citenamefont [1]{#1}%
\providecommand \href@noop [0]{\@secondoftwo}%
\providecommand \href [0]{\begingroup \@sanitize@url \@href}%
\providecommand \@href[1]{\@@startlink{#1}\@@href}%
\providecommand \@@href[1]{\endgroup#1\@@endlink}%
\providecommand \@sanitize@url [0]{\catcode `\\12\catcode `\$12\catcode
  `\&12\catcode `\#12\catcode `\^12\catcode `\_12\catcode `\%12\relax}%
\providecommand \@@startlink[1]{}%
\providecommand \@@endlink[0]{}%
\providecommand \url  [0]{\begingroup\@sanitize@url \@url }%
\providecommand \@url [1]{\endgroup\@href {#1}{\urlprefix }}%
\providecommand \urlprefix  [0]{URL }%
\providecommand \Eprint [0]{\href }%
\providecommand \doibase [0]{http://dx.doi.org/}%
\providecommand \selectlanguage [0]{\@gobble}%
\providecommand \bibinfo  [0]{\@secondoftwo}%
\providecommand \bibfield  [0]{\@secondoftwo}%
\providecommand \translation [1]{[#1]}%
\providecommand \BibitemOpen [0]{}%
\providecommand \bibitemStop [0]{}%
\providecommand \bibitemNoStop [0]{.\EOS\space}%
\providecommand \EOS [0]{\spacefactor3000\relax}%
\providecommand \BibitemShut  [1]{\csname bibitem#1\endcsname}%
\let\auto@bib@innerbib\@empty
\bibitem [{\citenamefont {Baran}\ \emph
  {et~al.}(2015{\natexlab{a}})\citenamefont {Baran}, \citenamefont {Palade},
  \citenamefont {Colonna}, \citenamefont {Di~Toro}, \citenamefont {Croitoru},\
  and\ \citenamefont {Nicolin}}]{PhysRevC.91.054303}%
  \BibitemOpen
  \bibfield  {author} {\bibinfo {author} {\bibfnamefont {V.}~\bibnamefont
  {Baran}}, \bibinfo {author} {\bibfnamefont {D.~I.}\ \bibnamefont {Palade}},
  \bibinfo {author} {\bibfnamefont {M.}~\bibnamefont {Colonna}}, \bibinfo
  {author} {\bibfnamefont {M.}~\bibnamefont {Di~Toro}}, \bibinfo {author}
  {\bibfnamefont {A.}~\bibnamefont {Croitoru}}, \ and\ \bibinfo {author}
  {\bibfnamefont {A.~I.}\ \bibnamefont {Nicolin}},\ }\href {\doibase
  10.1103/PhysRevC.91.054303} {\bibfield  {journal} {\bibinfo  {journal} {Phys.
  Rev. C}\ }\textbf {\bibinfo {volume} {91}},\ \bibinfo {pages} {054303}
  (\bibinfo {year} {2015}{\natexlab{a}})}\BibitemShut {NoStop}%
\bibitem [{\citenamefont {Baran}\ \emph
  {et~al.}(2015{\natexlab{b}})\citenamefont {Baran}, \citenamefont {Marciu},
  \citenamefont {Palade}, \citenamefont {Colonna}, \citenamefont {Di~Toro},
  \citenamefont {Nicolin},\ and\ \citenamefont {Zus}}]{baranromj}%
  \BibitemOpen
  \bibfield  {author} {\bibinfo {author} {\bibfnamefont {V.}~\bibnamefont
  {Baran}}, \bibinfo {author} {\bibfnamefont {M.}~\bibnamefont {Marciu}},
  \bibinfo {author} {\bibfnamefont {D.}~\bibnamefont {Palade}}, \bibinfo
  {author} {\bibfnamefont {M.}~\bibnamefont {Colonna}}, \bibinfo {author}
  {\bibfnamefont {M.}~\bibnamefont {Di~Toro}}, \bibinfo {author} {\bibfnamefont
  {A.}~\bibnamefont {Nicolin}}, \ and\ \bibinfo {author} {\bibfnamefont
  {R.}~\bibnamefont {Zus}},\ }\href
  {http://www.nipne.ro/rjp/2015_60_5-6/0727_0737.pdf} {\bibfield  {journal}
  {\bibinfo  {journal} {Rom. Journ. Phys.}\ }\textbf {\bibinfo {volume} {60}},\
  \bibinfo {pages} {727–737} (\bibinfo {year}
  {2015}{\natexlab{b}})}\BibitemShut {NoStop}%
\bibitem [{\citenamefont {Brown}\ and\ \citenamefont
  {Bolsterli}(1959)}]{brown1959dipole}%
  \BibitemOpen
  \bibfield  {author} {\bibinfo {author} {\bibfnamefont {G.~E.}\ \bibnamefont
  {Brown}}\ and\ \bibinfo {author} {\bibfnamefont {M.}~\bibnamefont
  {Bolsterli}},\ }\href {\doibase 10.1103/PhysRevLett.3.472} {\bibfield
  {journal} {\bibinfo  {journal} {Phys. Rev. Lett.}\ }\textbf {\bibinfo
  {volume} {3}},\ \bibinfo {pages} {472} (\bibinfo {year} {1959})}\BibitemShut
  {NoStop}%
\bibitem [{\citenamefont {Kresin}(1992)}]{kresin1992collective}%
  \BibitemOpen
  \bibfield  {author} {\bibinfo {author} {\bibfnamefont {V.~V.}\ \bibnamefont
  {Kresin}},\ }\href@noop {} {\bibfield  {journal} {\bibinfo  {journal}
  {Physics Reports}\ }\textbf {\bibinfo {volume} {220}},\ \bibinfo {pages} {1}
  (\bibinfo {year} {1992})}\BibitemShut {NoStop}%
\bibitem [{\citenamefont {de~Heer}(1993)}]{deheer}%
  \BibitemOpen
  \bibfield  {author} {\bibinfo {author} {\bibfnamefont {W.~A.}\ \bibnamefont
  {de~Heer}},\ }\href {\doibase 10.1103/RevModPhys.65.611} {\bibfield
  {journal} {\bibinfo  {journal} {Rev. Mod. Phys.}\ }\textbf {\bibinfo {volume}
  {65}},\ \bibinfo {pages} {611} (\bibinfo {year} {1993})}\BibitemShut
  {NoStop}%
\bibitem [{\citenamefont {Goldhaber}\ and\ \citenamefont
  {Teller}(1948)}]{goldhaber1948nuclear}%
  \BibitemOpen
  \bibfield  {author} {\bibinfo {author} {\bibfnamefont {M.}~\bibnamefont
  {Goldhaber}}\ and\ \bibinfo {author} {\bibfnamefont {E.}~\bibnamefont
  {Teller}},\ }\href {\doibase 10.1103/PhysRev.74.1046} {\bibfield  {journal}
  {\bibinfo  {journal} {Phys. Rev.}\ }\textbf {\bibinfo {volume} {74}},\
  \bibinfo {pages} {1046} (\bibinfo {year} {1948})}\BibitemShut {NoStop}%
\bibitem [{\citenamefont {Steinwedel}\ and\ \citenamefont
  {Jensen}(1950)}]{steinwedel1950z}%
  \BibitemOpen
  \bibfield  {author} {\bibinfo {author} {\bibfnamefont {H.}~\bibnamefont
  {Steinwedel}}\ and\ \bibinfo {author} {\bibfnamefont {H.}~\bibnamefont
  {Jensen}},\ }\href@noop {} {\bibfield  {journal} {\bibinfo  {journal} {Z.
  Naturforsch.}\ }\textbf {\bibinfo {volume} {5}},\ \bibinfo {pages} {413}
  (\bibinfo {year} {1950})}\BibitemShut {NoStop}%
\bibitem [{\citenamefont {Brack}(1993)}]{brack1993physics}%
  \BibitemOpen
  \bibfield  {author} {\bibinfo {author} {\bibfnamefont {M.}~\bibnamefont
  {Brack}},\ }\href {\doibase 10.1103/RevModPhys.65.677} {\bibfield  {journal}
  {\bibinfo  {journal} {Rev. Mod. Phys.}\ }\textbf {\bibinfo {volume} {65}},\
  \bibinfo {pages} {677} (\bibinfo {year} {1993})}\BibitemShut {NoStop}%
\bibitem [{\citenamefont {Van~Leeuwen}(2001)}]{van2001key}%
  \BibitemOpen
  \bibfield  {author} {\bibinfo {author} {\bibfnamefont {R.}~\bibnamefont
  {Van~Leeuwen}},\ }\href {\doibase 10.1142/S021797920100499X} {\bibfield
  {journal} {\bibinfo  {journal} {International Journal of Modern Physics B}\
  }\textbf {\bibinfo {volume} {15}},\ \bibinfo {pages} {1969} (\bibinfo {year}
  {2001})}\BibitemShut {NoStop}%
\bibitem [{\citenamefont {Calvayrac}\ \emph {et~al.}(1997)\citenamefont
  {Calvayrac}, \citenamefont {Reinhard},\ and\ \citenamefont
  {Suraud}}]{calvayrac1997spectral}%
  \BibitemOpen
  \bibfield  {author} {\bibinfo {author} {\bibfnamefont {F.}~\bibnamefont
  {Calvayrac}}, \bibinfo {author} {\bibfnamefont {P.}~\bibnamefont {Reinhard}},
  \ and\ \bibinfo {author} {\bibfnamefont {E.}~\bibnamefont {Suraud}},\ }\href
  {\doibase http://dx.doi.org/10.1006/aphy.1996.5654} {\bibfield  {journal}
  {\bibinfo  {journal} {Annals of Physics}\ }\textbf {\bibinfo {volume}
  {255}},\ \bibinfo {pages} {125 } (\bibinfo {year} {1997})}\BibitemShut
  {NoStop}%
\bibitem [{\citenamefont {Kohn}\ and\ \citenamefont
  {Sham}(1965)}]{Kohnsham1965}%
  \BibitemOpen
  \bibfield  {author} {\bibinfo {author} {\bibfnamefont {W.}~\bibnamefont
  {Kohn}}\ and\ \bibinfo {author} {\bibfnamefont {L.~J.}\ \bibnamefont
  {Sham}},\ }\href {\doibase 10.1103/PhysRev.140.A1133} {\bibfield  {journal}
  {\bibinfo  {journal} {Phys. Rev.}\ }\textbf {\bibinfo {volume} {140}},\
  \bibinfo {pages} {A1133} (\bibinfo {year} {1965})}\BibitemShut {NoStop}%
\bibitem [{\citenamefont {Lipparini}\ and\ \citenamefont
  {Stringari}(1991)}]{lipparini1991collective}%
  \BibitemOpen
  \bibfield  {author} {\bibinfo {author} {\bibfnamefont {E.}~\bibnamefont
  {Lipparini}}\ and\ \bibinfo {author} {\bibfnamefont {S.}~\bibnamefont
  {Stringari}},\ }\href@noop {} {\bibfield  {journal} {\bibinfo  {journal}
  {Zeitschrift f{\"u}r Physik D Atoms, Molecules and Clusters}\ }\textbf
  {\bibinfo {volume} {18}},\ \bibinfo {pages} {193} (\bibinfo {year}
  {1991})}\BibitemShut {NoStop}%
\bibitem [{\citenamefont {Reinhard}\ \emph {et~al.}(1990)\citenamefont
  {Reinhard}, \citenamefont {Brack},\ and\ \citenamefont
  {Genzken}}]{reinhard1990random}%
  \BibitemOpen
  \bibfield  {author} {\bibinfo {author} {\bibfnamefont {P.-G.}\ \bibnamefont
  {Reinhard}}, \bibinfo {author} {\bibfnamefont {M.}~\bibnamefont {Brack}}, \
  and\ \bibinfo {author} {\bibfnamefont {O.}~\bibnamefont {Genzken}},\
  }\href@noop {} {\bibfield  {journal} {\bibinfo  {journal} {Physical Review
  A}\ }\textbf {\bibinfo {volume} {41}},\ \bibinfo {pages} {5568} (\bibinfo
  {year} {1990})}\BibitemShut {NoStop}%
\bibitem [{\citenamefont {Nesterenko}\ \emph {et~al.}(1995)\citenamefont
  {Nesterenko}, \citenamefont {Kleinig},\ and\ \citenamefont
  {Gudkov}}]{nesterenko1995collectiveelambda}%
  \BibitemOpen
  \bibfield  {author} {\bibinfo {author} {\bibfnamefont {V.}~\bibnamefont
  {Nesterenko}}, \bibinfo {author} {\bibfnamefont {W.}~\bibnamefont {Kleinig}},
  \ and\ \bibinfo {author} {\bibfnamefont {V.}~\bibnamefont {Gudkov}},\
  }\href@noop {} {\bibfield  {journal} {\bibinfo  {journal} {Zeitschrift
  f{\"u}r Physik D Atoms,Molecules and Clusters}\ }\textbf {\bibinfo {volume}
  {34}},\ \bibinfo {pages} {271} (\bibinfo {year} {1995})}\BibitemShut
  {NoStop}%
\bibitem [{\citenamefont {Nesterenko}\ \emph {et~al.}(1997)\citenamefont
  {Nesterenko}, \citenamefont {Kleinig}, \citenamefont {Gudkov}, \citenamefont
  {Iudice},\ and\ \citenamefont {Kvasil}}]{nesterenko1997multipole}%
  \BibitemOpen
  \bibfield  {author} {\bibinfo {author} {\bibfnamefont {V.}~\bibnamefont
  {Nesterenko}}, \bibinfo {author} {\bibfnamefont {W.}~\bibnamefont {Kleinig}},
  \bibinfo {author} {\bibfnamefont {V.}~\bibnamefont {Gudkov}}, \bibinfo
  {author} {\bibfnamefont {N.~L.}\ \bibnamefont {Iudice}}, \ and\ \bibinfo
  {author} {\bibfnamefont {J.}~\bibnamefont {Kvasil}},\ }\href@noop {}
  {\bibfield  {journal} {\bibinfo  {journal} {Physical Review A}\ }\textbf
  {\bibinfo {volume} {56}},\ \bibinfo {pages} {607} (\bibinfo {year}
  {1997})}\BibitemShut {NoStop}%
\bibitem [{\citenamefont {Nesterenko}\ \emph {et~al.}(2002)\citenamefont
  {Nesterenko}, \citenamefont {Kvasil},\ and\ \citenamefont
  {Reinhard}}]{nest2002}%
  \BibitemOpen
  \bibfield  {author} {\bibinfo {author} {\bibfnamefont {V.~O.}\ \bibnamefont
  {Nesterenko}}, \bibinfo {author} {\bibfnamefont {J.}~\bibnamefont {Kvasil}},
  \ and\ \bibinfo {author} {\bibfnamefont {P.-G.}\ \bibnamefont {Reinhard}},\
  }\href {\doibase 10.1103/PhysRevC.66.044307} {\bibfield  {journal} {\bibinfo
  {journal} {Phys. Rev. C}\ }\textbf {\bibinfo {volume} {66}},\ \bibinfo
  {pages} {044307} (\bibinfo {year} {2002})}\BibitemShut {NoStop}%
\bibitem [{\citenamefont {Runge}\ and\ \citenamefont
  {Gross}(1984)}]{runge1984time}%
  \BibitemOpen
  \bibfield  {author} {\bibinfo {author} {\bibfnamefont {E.}~\bibnamefont
  {Runge}}\ and\ \bibinfo {author} {\bibfnamefont {E.~K.~U.}\ \bibnamefont
  {Gross}},\ }\href {\doibase 10.1103/PhysRevLett.52.997} {\bibfield  {journal}
  {\bibinfo  {journal} {Phys. Rev. Lett.}\ }\textbf {\bibinfo {volume} {52}},\
  \bibinfo {pages} {997} (\bibinfo {year} {1984})}\BibitemShut {NoStop}%
\bibitem [{\citenamefont {Bauernschmitt}\ and\ \citenamefont
  {Ahlrichs}(1996)}]{bauernschmitt1996treatment}%
  \BibitemOpen
  \bibfield  {author} {\bibinfo {author} {\bibfnamefont {R.}~\bibnamefont
  {Bauernschmitt}}\ and\ \bibinfo {author} {\bibfnamefont {R.}~\bibnamefont
  {Ahlrichs}},\ }\href@noop {} {\bibfield  {journal} {\bibinfo  {journal}
  {Chemical Physics Letters}\ }\textbf {\bibinfo {volume} {256}},\ \bibinfo
  {pages} {454} (\bibinfo {year} {1996})}\BibitemShut {NoStop}%
\bibitem [{\citenamefont {Muta}\ \emph {et~al.}(2002)\citenamefont {Muta},
  \citenamefont {Iwata}, \citenamefont {Hashimoto},\ and\ \citenamefont
  {Yabana}}]{muta2002solving}%
  \BibitemOpen
  \bibfield  {author} {\bibinfo {author} {\bibfnamefont {A.}~\bibnamefont
  {Muta}}, \bibinfo {author} {\bibfnamefont {J.-I.}\ \bibnamefont {Iwata}},
  \bibinfo {author} {\bibfnamefont {Y.}~\bibnamefont {Hashimoto}}, \ and\
  \bibinfo {author} {\bibfnamefont {K.}~\bibnamefont {Yabana}},\ }\href@noop {}
  {\bibfield  {journal} {\bibinfo  {journal} {Progress of theoretical physics}\
  }\textbf {\bibinfo {volume} {108}},\ \bibinfo {pages} {1065} (\bibinfo {year}
  {2002})}\BibitemShut {NoStop}%
\bibitem [{\citenamefont {Yabana}\ \emph {et~al.}(2006)\citenamefont {Yabana},
  \citenamefont {Nakatsukasa}, \citenamefont {Iwata},\ and\ \citenamefont
  {Bertsch}}]{yabana2006real}%
  \BibitemOpen
  \bibfield  {author} {\bibinfo {author} {\bibfnamefont {K.}~\bibnamefont
  {Yabana}}, \bibinfo {author} {\bibfnamefont {T.}~\bibnamefont {Nakatsukasa}},
  \bibinfo {author} {\bibfnamefont {J.}~\bibnamefont {Iwata}}, \ and\ \bibinfo
  {author} {\bibfnamefont {G.}~\bibnamefont {Bertsch}},\ }\href@noop {}
  {\bibfield  {journal} {\bibinfo  {journal} {Physica Status Solidi B Basic
  Research}\ }\textbf {\bibinfo {volume} {243}},\ \bibinfo {pages} {1121}
  (\bibinfo {year} {2006})}\BibitemShut {NoStop}%
\bibitem [{\citenamefont {Reinhard}\ and\ \citenamefont
  {Gambhir}(1992)}]{reinhard1992rpa}%
  \BibitemOpen
  \bibfield  {author} {\bibinfo {author} {\bibfnamefont {P.-G.}\ \bibnamefont
  {Reinhard}}\ and\ \bibinfo {author} {\bibfnamefont {Y.}~\bibnamefont
  {Gambhir}},\ }\href@noop {} {\bibfield  {journal} {\bibinfo  {journal}
  {Annalen der Physik}\ }\textbf {\bibinfo {volume} {504}},\ \bibinfo {pages}
  {598} (\bibinfo {year} {1992})}\BibitemShut {NoStop}%
\bibitem [{\citenamefont {Reinhard}\ \emph {et~al.}(1996)\citenamefont
  {Reinhard}, \citenamefont {Genzken},\ and\ \citenamefont
  {Brack}}]{reinhard1996sum}%
  \BibitemOpen
  \bibfield  {author} {\bibinfo {author} {\bibfnamefont {P.-G.}\ \bibnamefont
  {Reinhard}}, \bibinfo {author} {\bibfnamefont {O.}~\bibnamefont {Genzken}}, \
  and\ \bibinfo {author} {\bibfnamefont {M.}~\bibnamefont {Brack}},\
  }\href@noop {} {\bibfield  {journal} {\bibinfo  {journal} {Annalen der
  Physik}\ }\textbf {\bibinfo {volume} {508}},\ \bibinfo {pages} {576}
  (\bibinfo {year} {1996})}\BibitemShut {NoStop}%
\bibitem [{\citenamefont {Wang}(1999)}]{TKR}%
  \BibitemOpen
  \bibfield  {author} {\bibinfo {author} {\bibfnamefont {S.}~\bibnamefont
  {Wang}},\ }\href {\doibase 10.1103/PhysRevA.60.262} {\bibfield  {journal}
  {\bibinfo  {journal} {Phys. Rev. A}\ }\textbf {\bibinfo {volume} {60}},\
  \bibinfo {pages} {262} (\bibinfo {year} {1999})}\BibitemShut {NoStop}%
\bibitem [{\citenamefont {Gunnarsson}\ and\ \citenamefont
  {Lundqvist}(1976)}]{gunarsson}%
  \BibitemOpen
  \bibfield  {author} {\bibinfo {author} {\bibfnamefont {O.}~\bibnamefont
  {Gunnarsson}}\ and\ \bibinfo {author} {\bibfnamefont {B.~I.}\ \bibnamefont
  {Lundqvist}},\ }\href {\doibase 10.1103/PhysRevB.13.4274} {\bibfield
  {journal} {\bibinfo  {journal} {Phys. Rev. B}\ }\textbf {\bibinfo {volume}
  {13}},\ \bibinfo {pages} {4274} (\bibinfo {year} {1976})}\BibitemShut
  {NoStop}%
\bibitem [{\citenamefont {Calvayrac}\ \emph {et~al.}(2000)\citenamefont
  {Calvayrac}, \citenamefont {Reinhard}, \citenamefont {Suraud},\ and\
  \citenamefont {Ullrich}}]{Calvayrac2000493}%
  \BibitemOpen
  \bibfield  {author} {\bibinfo {author} {\bibfnamefont {F.}~\bibnamefont
  {Calvayrac}}, \bibinfo {author} {\bibfnamefont {P.-G.}\ \bibnamefont
  {Reinhard}}, \bibinfo {author} {\bibfnamefont {E.}~\bibnamefont {Suraud}}, \
  and\ \bibinfo {author} {\bibfnamefont {C.}~\bibnamefont {Ullrich}},\ }\href
  {\doibase http://dx.doi.org/10.1016/S0370-1573(00)00043-0} {\bibfield
  {journal} {\bibinfo  {journal} {Physics Reports}\ }\textbf {\bibinfo {volume}
  {337}},\ \bibinfo {pages} {493 } (\bibinfo {year} {2000})}\BibitemShut
  {NoStop}%
\bibitem [{\citenamefont {Yabana}\ and\ \citenamefont
  {Bertsch}(1996)}]{YABANAREAL}%
  \BibitemOpen
  \bibfield  {author} {\bibinfo {author} {\bibfnamefont {K.}~\bibnamefont
  {Yabana}}\ and\ \bibinfo {author} {\bibfnamefont {G.~F.}\ \bibnamefont
  {Bertsch}},\ }\href {\doibase 10.1103/PhysRevB.54.4484} {\bibfield  {journal}
  {\bibinfo  {journal} {Phys. Rev. B}\ }\textbf {\bibinfo {volume} {54}},\
  \bibinfo {pages} {4484} (\bibinfo {year} {1996})}\BibitemShut {NoStop}%
\bibitem [{\citenamefont {Yannouleas}\ \emph {et~al.}(1993)\citenamefont
  {Yannouleas}, \citenamefont {Vigezzi},\ and\ \citenamefont
  {Broglia}}]{yannouleas1993evolution}%
  \BibitemOpen
  \bibfield  {author} {\bibinfo {author} {\bibfnamefont {C.}~\bibnamefont
  {Yannouleas}}, \bibinfo {author} {\bibfnamefont {E.}~\bibnamefont {Vigezzi}},
  \ and\ \bibinfo {author} {\bibfnamefont {R.~A.}\ \bibnamefont {Broglia}},\
  }\href@noop {} {\bibfield  {journal} {\bibinfo  {journal} {Physical Review
  B}\ }\textbf {\bibinfo {volume} {47}},\ \bibinfo {pages} {9849} (\bibinfo
  {year} {1993})}\BibitemShut {NoStop}%
\bibitem [{\citenamefont {Rowe}(1970)}]{Rowe_1970}%
  \BibitemOpen
  \bibfield  {author} {\bibinfo {author} {\bibfnamefont {D.}~\bibnamefont
  {Rowe}},\ }\href@noop {} {\emph {\bibinfo {title} {NUCLEAR COLLECTIVE MOTION.
  MODELS AND THEORY.}}}\ (\bibinfo  {publisher} {1951},\ \bibinfo {year}
  {1970})\BibitemShut {NoStop}%
\bibitem [{\citenamefont {Tikhonov}\ \emph {et~al.}(2001)\citenamefont
  {Tikhonov}, \citenamefont {Kasperovich}, \citenamefont {Wong},\ and\
  \citenamefont {Kresin}}]{tikhonovstatic}%
  \BibitemOpen
  \bibfield  {author} {\bibinfo {author} {\bibfnamefont {G.}~\bibnamefont
  {Tikhonov}}, \bibinfo {author} {\bibfnamefont {V.}~\bibnamefont
  {Kasperovich}}, \bibinfo {author} {\bibfnamefont {K.}~\bibnamefont {Wong}}, \
  and\ \bibinfo {author} {\bibfnamefont {V.~V.}\ \bibnamefont {Kresin}},\
  }\href {\doibase 10.1103/PhysRevA.64.063202} {\bibfield  {journal} {\bibinfo
  {journal} {Phys. Rev. A}\ }\textbf {\bibinfo {volume} {64}},\ \bibinfo
  {pages} {063202} (\bibinfo {year} {2001})}\BibitemShut {NoStop}%
\bibitem [{\citenamefont {Xia}\ \emph {et~al.}(2009)\citenamefont {Xia},
  \citenamefont {Yin},\ and\ \citenamefont {Kresin}}]{XiaKresin2009}%
  \BibitemOpen
  \bibfield  {author} {\bibinfo {author} {\bibfnamefont {C.}~\bibnamefont
  {Xia}}, \bibinfo {author} {\bibfnamefont {C.}~\bibnamefont {Yin}}, \ and\
  \bibinfo {author} {\bibfnamefont {V.~V.}\ \bibnamefont {Kresin}},\ }\href
  {\doibase 10.1103/PhysRevLett.102.156802} {\bibfield  {journal} {\bibinfo
  {journal} {Phys. Rev. Lett.}\ }\textbf {\bibinfo {volume} {102}},\ \bibinfo
  {pages} {156802} (\bibinfo {year} {2009})}\BibitemShut {NoStop}%
\bibitem [{\citenamefont {Schmidt}\ and\ \citenamefont
  {Haberland}(1999)}]{goodexp}%
  \BibitemOpen
  \bibfield  {author} {\bibinfo {author} {\bibfnamefont {M.}~\bibnamefont
  {Schmidt}}\ and\ \bibinfo {author} {\bibfnamefont {H.}~\bibnamefont
  {Haberland}},\ }\href {\doibase 10.1007/s100530050290} {\bibfield  {journal}
  {\bibinfo  {journal} {Eur. Phys. J. D}\ }\textbf {\bibinfo {volume} {6}},\
  \bibinfo {pages} {109} (\bibinfo {year} {1999})}\BibitemShut {NoStop}%
\bibitem [{\citenamefont {Kroto}\ \emph {et~al.}(1985)\citenamefont {Kroto},
  \citenamefont {Heath}, \citenamefont {O'Brien}, \citenamefont {Curl},
  \citenamefont {Smalley} \emph {et~al.}}]{kroto1985c}%
  \BibitemOpen
  \bibfield  {author} {\bibinfo {author} {\bibfnamefont {H.~W.}\ \bibnamefont
  {Kroto}}, \bibinfo {author} {\bibfnamefont {J.~R.}\ \bibnamefont {Heath}},
  \bibinfo {author} {\bibfnamefont {S.~C.}\ \bibnamefont {O'Brien}}, \bibinfo
  {author} {\bibfnamefont {R.~F.}\ \bibnamefont {Curl}}, \bibinfo {author}
  {\bibfnamefont {R.~E.}\ \bibnamefont {Smalley}},  \emph {et~al.},\
  }\href@noop {} {\bibfield  {journal} {\bibinfo  {journal} {Nature}\ }\textbf
  {\bibinfo {volume} {318}},\ \bibinfo {pages} {162} (\bibinfo {year}
  {1985})}\BibitemShut {NoStop}%
\bibitem [{\citenamefont {Scully}\ \emph {et~al.}(2005)\citenamefont {Scully},
  \citenamefont {Emmons}, \citenamefont {Gharaibeh}, \citenamefont {Phaneuf},
  \citenamefont {Kilcoyne}, \citenamefont {Schlachter}, \citenamefont
  {Schippers}, \citenamefont {M\"uller}, \citenamefont {Chakraborty},
  \citenamefont {Madjet},\ and\ \citenamefont {Rost}}]{scully2005}%
  \BibitemOpen
  \bibfield  {author} {\bibinfo {author} {\bibfnamefont {S.~W.~J.}\
  \bibnamefont {Scully}}, \bibinfo {author} {\bibfnamefont {E.~D.}\
  \bibnamefont {Emmons}}, \bibinfo {author} {\bibfnamefont {M.~F.}\
  \bibnamefont {Gharaibeh}}, \bibinfo {author} {\bibfnamefont {R.~A.}\
  \bibnamefont {Phaneuf}}, \bibinfo {author} {\bibfnamefont {A.~L.~D.}\
  \bibnamefont {Kilcoyne}}, \bibinfo {author} {\bibfnamefont {A.~S.}\
  \bibnamefont {Schlachter}}, \bibinfo {author} {\bibfnamefont
  {S.}~\bibnamefont {Schippers}}, \bibinfo {author} {\bibfnamefont
  {A.}~\bibnamefont {M\"uller}}, \bibinfo {author} {\bibfnamefont {H.~S.}\
  \bibnamefont {Chakraborty}}, \bibinfo {author} {\bibfnamefont {M.~E.}\
  \bibnamefont {Madjet}}, \ and\ \bibinfo {author} {\bibfnamefont {J.~M.}\
  \bibnamefont {Rost}},\ }\href {\doibase 10.1103/PhysRevLett.94.065503}
  {\bibfield  {journal} {\bibinfo  {journal} {Phys. Rev. Lett.}\ }\textbf
  {\bibinfo {volume} {94}},\ \bibinfo {pages} {065503} (\bibinfo {year}
  {2005})}\BibitemShut {NoStop}%
\bibitem [{\citenamefont {Verkhovtsev}\ \emph {et~al.}(2012)\citenamefont
  {Verkhovtsev}, \citenamefont {Korol}, \citenamefont {Solov'yov},
  \citenamefont {Bolognesi}, \citenamefont {Ruocco},\ and\ \citenamefont
  {Avaldi}}]{korolees}%
  \BibitemOpen
  \bibfield  {author} {\bibinfo {author} {\bibfnamefont {A.~V.}\ \bibnamefont
  {Verkhovtsev}}, \bibinfo {author} {\bibfnamefont {A.~V.}\ \bibnamefont
  {Korol}}, \bibinfo {author} {\bibfnamefont {A.~V.}\ \bibnamefont
  {Solov'yov}}, \bibinfo {author} {\bibfnamefont {P.}~\bibnamefont
  {Bolognesi}}, \bibinfo {author} {\bibfnamefont {A.}~\bibnamefont {Ruocco}}, \
  and\ \bibinfo {author} {\bibfnamefont {L.}~\bibnamefont {Avaldi}},\ }\href
  {http://stacks.iop.org/0953-4075/45/i=14/a=141002} {\bibfield  {journal}
  {\bibinfo  {journal} {Journal of Physics B: Atomic, Molecular and Optical
  Physics}\ }\textbf {\bibinfo {volume} {45}},\ \bibinfo {pages} {141002}
  (\bibinfo {year} {2012})}\BibitemShut {NoStop}%
\bibitem [{\citenamefont {Bertsch}\ \emph {et~al.}(1991)\citenamefont
  {Bertsch}, \citenamefont {Bulgac}, \citenamefont {Tom\'anek},\ and\
  \citenamefont {Wang}}]{bertschcollective1991}%
  \BibitemOpen
  \bibfield  {author} {\bibinfo {author} {\bibfnamefont {G.~F.}\ \bibnamefont
  {Bertsch}}, \bibinfo {author} {\bibfnamefont {A.}~\bibnamefont {Bulgac}},
  \bibinfo {author} {\bibfnamefont {D.}~\bibnamefont {Tom\'anek}}, \ and\
  \bibinfo {author} {\bibfnamefont {Y.}~\bibnamefont {Wang}},\ }\href {\doibase
  10.1103/PhysRevLett.67.2690} {\bibfield  {journal} {\bibinfo  {journal}
  {Phys. Rev. Lett.}\ }\textbf {\bibinfo {volume} {67}},\ \bibinfo {pages}
  {2690} (\bibinfo {year} {1991})}\BibitemShut {NoStop}%
\bibitem [{\citenamefont {Palade}\ and\ \citenamefont {Baran}()}]{palade2015}%
  \BibitemOpen
  \bibfield  {author} {\bibinfo {author} {\bibfnamefont {D.~I.}\ \bibnamefont
  {Palade}}\ and\ \bibinfo {author} {\bibfnamefont {V.}~\bibnamefont {Baran}},\
  }\href@noop {} {\bibfield  {journal} {\bibinfo  {journal} {Journal of Physics
  B: Atomic, Molecular and Optical Physics}\ }\textbf {\bibinfo {volume} {In
  press}}}\BibitemShut {NoStop}%
\bibitem [{\citenamefont {Madjet}\ \emph {et~al.}(2008)\citenamefont {Madjet},
  \citenamefont {Chakraborty}, \citenamefont {Rost},\ and\ \citenamefont
  {Manson}}]{madjet2008}%
  \BibitemOpen
  \bibfield  {author} {\bibinfo {author} {\bibfnamefont {M.~E.}\ \bibnamefont
  {Madjet}}, \bibinfo {author} {\bibfnamefont {H.~S.}\ \bibnamefont
  {Chakraborty}}, \bibinfo {author} {\bibfnamefont {J.~M.}\ \bibnamefont
  {Rost}}, \ and\ \bibinfo {author} {\bibfnamefont {S.~T.}\ \bibnamefont
  {Manson}},\ }\href {http://stacks.iop.org/0953-4075/41/i=10/a=105101}
  {\bibfield  {journal} {\bibinfo  {journal} {Journal of Physics B: Atomic,
  Molecular and Optical Physics}\ }\textbf {\bibinfo {volume} {41}},\ \bibinfo
  {pages} {105101} (\bibinfo {year} {2008})}\BibitemShut {NoStop}%
\bibitem [{\citenamefont {Puska}\ and\ \citenamefont
  {Nieminen}(1993)}]{pusk1993photo}%
  \BibitemOpen
  \bibfield  {author} {\bibinfo {author} {\bibfnamefont {M.~J.}\ \bibnamefont
  {Puska}}\ and\ \bibinfo {author} {\bibfnamefont {R.~M.}\ \bibnamefont
  {Nieminen}},\ }\href {\doibase 10.1103/PhysRevA.47.1181} {\bibfield
  {journal} {\bibinfo  {journal} {Phys. Rev. A}\ }\textbf {\bibinfo {volume}
  {47}},\ \bibinfo {pages} {1181} (\bibinfo {year} {1993})}\BibitemShut
  {NoStop}%
\bibitem [{\citenamefont {Gueorguiev}\ \emph {et~al.}(2004)\citenamefont
  {Gueorguiev}, \citenamefont {Pacheco},\ and\ \citenamefont
  {Tom\'anek}}]{C60polar}%
  \BibitemOpen
  \bibfield  {author} {\bibinfo {author} {\bibfnamefont {G.~K.}\ \bibnamefont
  {Gueorguiev}}, \bibinfo {author} {\bibfnamefont {J.~M.}\ \bibnamefont
  {Pacheco}}, \ and\ \bibinfo {author} {\bibfnamefont {D.}~\bibnamefont
  {Tom\'anek}},\ }\href {\doibase 10.1103/PhysRevLett.92.215501} {\bibfield
  {journal} {\bibinfo  {journal} {Phys. Rev. Lett.}\ }\textbf {\bibinfo
  {volume} {92}},\ \bibinfo {pages} {215501} (\bibinfo {year}
  {2004})}\BibitemShut {NoStop}%
\bibitem [{\citenamefont {Kafle}\ \emph {et~al.}(2008)\citenamefont {Kafle},
  \citenamefont {Katayanagi}, \citenamefont {Prodhan}, \citenamefont {Yagi},
  \citenamefont {Huang},\ and\ \citenamefont {Mitsuke}}]{KafleC60}%
  \BibitemOpen
  \bibfield  {author} {\bibinfo {author} {\bibfnamefont {B.~P.}\ \bibnamefont
  {Kafle}}, \bibinfo {author} {\bibfnamefont {H.}~\bibnamefont {Katayanagi}},
  \bibinfo {author} {\bibfnamefont {M.~S.~I.}\ \bibnamefont {Prodhan}},
  \bibinfo {author} {\bibfnamefont {H.}~\bibnamefont {Yagi}}, \bibinfo {author}
  {\bibfnamefont {C.}~\bibnamefont {Huang}}, \ and\ \bibinfo {author}
  {\bibfnamefont {K.}~\bibnamefont {Mitsuke}},\ }\href {\doibase
  10.1143/JPSJ.77.014302} {\bibfield  {journal} {\bibinfo  {journal} {Journal
  of the Physical Society of Japan}\ }\textbf {\bibinfo {volume} {77}},\
  \bibinfo {pages} {014302} (\bibinfo {year} {2008})}\BibitemShut {NoStop}%
\end{thebibliography}%


%

\end{document}